\documentclass[%
reprint,
showpacs,preprintnumbers,
showkeys,
amsmath,amssymb,
aps,
prl,
floatfix,superscriptaddress
]{revtex4-1}

\usepackage{pdfpages}
\usepackage{pgffor}
\usepackage[english]{babel}
\usepackage[utf8]{inputenc}
\usepackage{graphicx}
\usepackage{hyperref}
\usepackage{xcolor}
\usepackage{amsmath}
\usepackage{amsfonts}
\usepackage{amssymb}
\usepackage{dsfont}

\makeatletter
\AtBeginDocument{\let\LS@rot\@undefined}
\makeatother

\definecolor{refColor}{HTML}{EA00F2}
\definecolor{figColor}{HTML}{008DF2}
\definecolor{urlColor}{HTML}{00AEF2}

\hypersetup{
    unicode=false,                 % non-Latin characters in Acrobat’s bookmarks
    pdftoolbar=true,               % show Acrobat’s toolbar?
    pdfmenubar=true,               % show Acrobat’s menu?
    pdffitwindow=false,            % window fit to page when opened
    pdfstartview={FitH},           % fits the width of the page to the window
    pdftitle={},                   % title
    pdfsubject={Quantum physics},  % subject of the document
    pdfkeywords={},                % list of keywords
    pdfnewwindow=true,             % links in new window
    colorlinks=true,               % false: boxed links; true: colored links
    linkcolor=figColor,            % color of internal links
    citecolor=refColor,            % color of links to bibliography
    filecolor=magenta,             % color of file links
    urlcolor=urlColor              % color of external links
}

\newcommand{\rs}[1]{\scriptscriptstyle \mathrm{ #1 }}

\newcommand{\bra}[1]{\left\langle #1\right|}             %% bra
\newcommand{\ket}[1]{\left| #1\right\rangle}              %% ket
\newcommand{\braket}[2]{\left\langle #1\middle|#2\right\rangle}              %% scalar product
\begin{document}

\title{Emergent Universal Dynamics for an Atomic Cloud Coupled to an Optical Waveguide}

\author{Jan Kumlin}
\email{kumlin@itp3.uni-stuttgart.de}
\affiliation{Institute for Theoretical Physics III and Center for Integrated Quantum Science and Technology, 
  University of Stuttgart, 70550 Stuttgart, Germany}
\author{Sebastian Hofferberth}
\affiliation{Department of Physics, Chemistry and Pharmacy, Physics@SDU, University of Southern Denmark, 5320 Odense, Denmark}
\author{Hans Peter B\"uchler}
%\email{buechler@itp3.uni-stuttgart.de}
\affiliation{Institute for Theoretical Physics III, 
  University of Stuttgart, 70550 Stuttgart, Germany}

\date{\today}

% PACS

% Keywords
%\keywords{}

\begin{abstract}
  We study the dynamics of a single collective excitation in a cold ensemble of atoms coupled to a one-dimensional waveguide. 
  The coupling between the atoms and the photonic modes provides a coherent and a dissipative dynamics
  for this collective excitation. While the dissipative part accounts for the collectively enhanced and directed emission of photons, we find a remarkable
  universal dynamics for increasing atom numbers exhibiting several revivals under the coherent part. While this phenomenon provides a limit on the intrinsic dephasing for such a collective excitation,
  a setup is presented, where the universal dynamics can be explored. 
\end{abstract}

\maketitle

The collective interaction between an ensemble of emitters and  photons is at the heart of many fascinating phenomena in quantum 
optics \cite{Dicke54,Gross1982}.  For a single coherent excitation of such an ensemble, the direction and the rate of spontaneous 
emission are strongly modified and can either be enhanced or suppressed, which has recently been experimentally observed \cite{Dudin2012,Guerin2016}. 
For these effects to be observable, it is crucial that the coherence  between the atoms within the ensemble is maintained.  While the influence of the thermal 
motion of atoms has been investigated \cite{Bromley2016},  an ensemble of atoms with a single excitation also exhibits an interaction induced by the virtual 
exchange of photons  \cite{Lehmberg1970}, which might provide  an intrinsic dephasing inherent to any ensemble of emitters. In this Letter, we study within 
a microscopic analysis whether there is a fundamental limit on this intrinsic dephasing.

Signatures of the coherent interaction by a virtual exchange of photons in an ensemble of atoms with several excitations have been discussed in terms of a collective Lamb shift 
\cite{Lehmberg1970, Friedberg1973}, and they have been observed in various physical systems ranging from an ensemble of nuclei \cite{Rohlsberger2010} 
over solid-state systems \cite{Scheibner2007, Tighineanu2016} to ions \cite{Meir2014} and atoms \cite{Pellegrino2014, Jennewein2016}.  On the theoretical side, 
recent research has focused on the understanding of the transmission of photons and the appearance of correlations in 
one-dimensional waveguides \cite{Pichler2015,Shi2015, Ruostekoski2016, Goban2014, Solano2017,Lodahl2017}, as well as the appearance of 
superradiance and collective Lamb shift in the single-excitation manifold \cite{Scully2006,Mazets2007,Svidzinsky2008,Scully2009,Javanainen2014,Garcia2017}.
 In order to guarantee a single excitation in an ensemble of
scatterers, the notion of a superatom has emerged as a powerful concept, where a strong interaction between the excited states restricts 
suppressed multiple excitations in the ensemble and is conveniently realized with Rydberg atoms  \cite{Jaksch2000,Lukin2001,Saffman2010,Ebert2014,Labuhn2016,Dudin2012a,Zeiher2015,Paris-Mandoki2017}.

 \begin{figure}[h]
\includegraphics[width=0.4\textwidth]{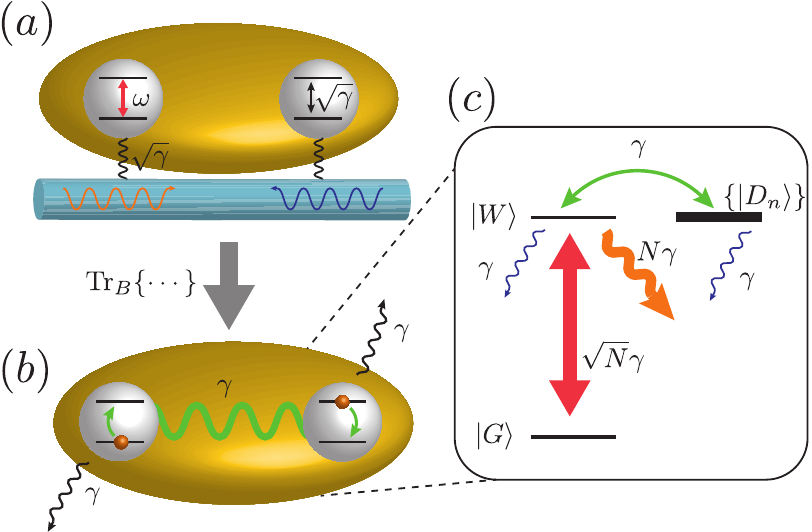}
\caption{(a) Two-level atoms coupled to a one-dimensional waveguide with left- and right-moving modes. (b) After integrating out the photonic degrees of freedom, the system exhibits spontaneous emission and an infinite-ranged exchange interaction between the atoms. (c) In the presence of a blockade mechanism, the  superatom state $\ket{W}$ is collectively coupled to the ground state with the coupling strength $\sqrt{N\gamma}$ giving rise to an enhanced spontaneous emission $N\gamma$ into the forward direction, while the coherent exchange interaction leads to a coupling between this bright state and the manifold of dark states. \label{fig:setup}}
\end{figure}

Here, we study the influence of the virtual exchange of photons on the properties of such a single collective excitation, focusing on a setup described by an ensemble of 
atoms coupled to a one-dimensional waveguide, see Fig.~\ref{fig:setup}. Based on a microscopic analysis, the time evolution of 
the collective excited state is governed by two competing terms: first, the spontaneous and strongly directed 
emission into the waveguide, and second an intrinsic coherent exchange interaction. Remarkably, we find that 
the coherent part gives rise  to  a universal dynamics of the collective excitation for  increasing particle numbers, 
and it exhibits several revivals and eventually saturates at a universal value. While this phenomenon provides an intrinsic 
limit on the dephasing in a superatom, we also present a setup, where the universal dynamics can be explored. 

Each atom is well-described by a two-level system with the ground state $\ket{g}$ and the excited state $\ket{e}$ (see Fig.~\ref{fig:setup}). The optical transition 
frequency between the two states is given by $\omega_0 = 2 \pi c / \lambda$, with the wave length $\lambda$ and the corresponding wave 
vector $k = 2\pi / \lambda$. In the following, we describe the two states of an atom at position $x$ by the field operators $\psi_g(x)$ and 
$\psi_e(x)$ for the ground and excited state, respectively. Then, the initial state with all $N$ atoms in the ground state takes the form 
$\ket{G} = \sqrt{1/ N!} \prod_{i = 1}^N \psi^\dagger_g(x_i) \ket{0}$. The atomic positions $x_i$ are randomly distributed with a distribution 
function giving rise to the averaged  atomic ground state density $n(x) = \langle G|  \psi^\dagger_g(x) \psi_g(x)|G \rangle_{\rs{dis}}$; here, $\langle \cdots \rangle_{\rs{dis}}$ 
denotes the ensemble average over many experimental realizations. Furthermore, we introduce the operators $S^+(x) = \psi^\dagger_e(x) \psi_g(x)$ 
creating an excitation from the ground state to the excited state and $S^-(x) = \psi^\dagger_g(x) \psi_e(x)$ for a transition from the excited 
state to the ground state. These operators satisfy the relation
\begin{equation}
[S^+(x), S^-(y)] = \delta(x-y) \left[\hat{n}_g(x) -\hat{ n}_e(x) \right]
\end{equation}
with $\hat{n}_\nu(x) = \psi^\dagger_\nu(x) \psi_\nu(x)$ for $\nu \in \{g,e\}$. Then, the microscopic Hamiltonian describing the coupling of the atoms to a one-dimensional waveguide within the rotating-wave approximation takes the form
\begin{align}
H &= \int \frac{dq}{2\pi} \hbar \omega_q a^\dagger_q a_q + \hbar \omega_0 \int dx \, \psi^\dagger_e(x) \psi_e(x) \nonumber \\
&\quad - \hbar \sqrt{\gamma} \int dx \, \left[\mathcal{E}^\dagger(x)  S^-(x)  + S^+(x) \mathcal{E}(x)  \right] \, ,
\label{eq:Hamiltonian_1D}
\end{align}
where $\sqrt{\gamma}$ characterizes the effective mode coupling  giving rise to 
the rate  $\gamma$ for spontaneous emission of a left- or right-moving photon in the waveguide
  \cite{Goban2014,Shi2015,Ruostekoski2016,Pichler2015,Solano2017,Lodahl2017}.
Furthermore, the electric field operator within the waveguide reduces to
\begin{equation}
\mathcal{E}^\dagger(x) = - i \sqrt{c}\, \int \frac{dq}{2\pi}  \, a_q^\dagger e^{-iqx} \, .
\end{equation}
The bosonic operators $a^\dagger_q$ account for the creation of a waveguide mode with momentum $q$, while  $\omega_q = c |q|$  %$\omega_q$ 
denotes the dispersion relation for the relevant photon modes.

Starting from the microscopic Hamiltonian (\ref{eq:Hamiltonian_1D}) and integrating out the electric field, the effective dynamics for the atoms alone is governed by a master equation \cite{Lehmberg1970,Pichler2015,sup}
and takes the form %\cite{Pichler2015, Shi2015}
\begin{equation}
\partial_t \rho = -\frac{i}{\hbar} [H_s, \rho] + \mathcal{D}_{F}[\rho] + \mathcal{D}_{ B}[\rho] \, . 
\end{equation}
The first term describes a coherent interaction between the atoms by the exchange of virtual photons,
\begin{equation}
H_s = \hbar \gamma \int dx \, dy \, \sin(k \vert x- y \vert) \, S^+(x) S^-(y) \, . 
\label{eq:Hs}
\end{equation}
The term $\mathcal{D}_{F}$ ($\mathcal{D}_{ B}$) describes the spontaneous emission of a photon in the forward (backward) propagating mode, respectively. 

In the following, the main analysis focuses on the superatom state
\begin{equation}
\ket{W} = \frac{1}{\sqrt{N}} \int dx \, e^{ikx} S^+(x) \ket{G}% \equiv \frac{1}{\sqrt{N}} c_+^\dagger \ket{G} 
\end{equation}
which couples to the incoming light field with the collectively enhanced coupling strength $\sqrt{N \gamma}$. In addition, there are $N-1$ "dark" states 
%\begin{equation}
$\ket{D_n} = \int dx \, D_n(x) S^+(x) \ket{G}$,
%\end{equation}
with the wave functions $D_n(x)$ defined by the orthogonality conditions $\braket{W}{D_n} = 0$ and $\braket{D_m}{D_n} = \delta_{nm}$.

\begin{figure}[h]
\includegraphics[width=0.5\textwidth]{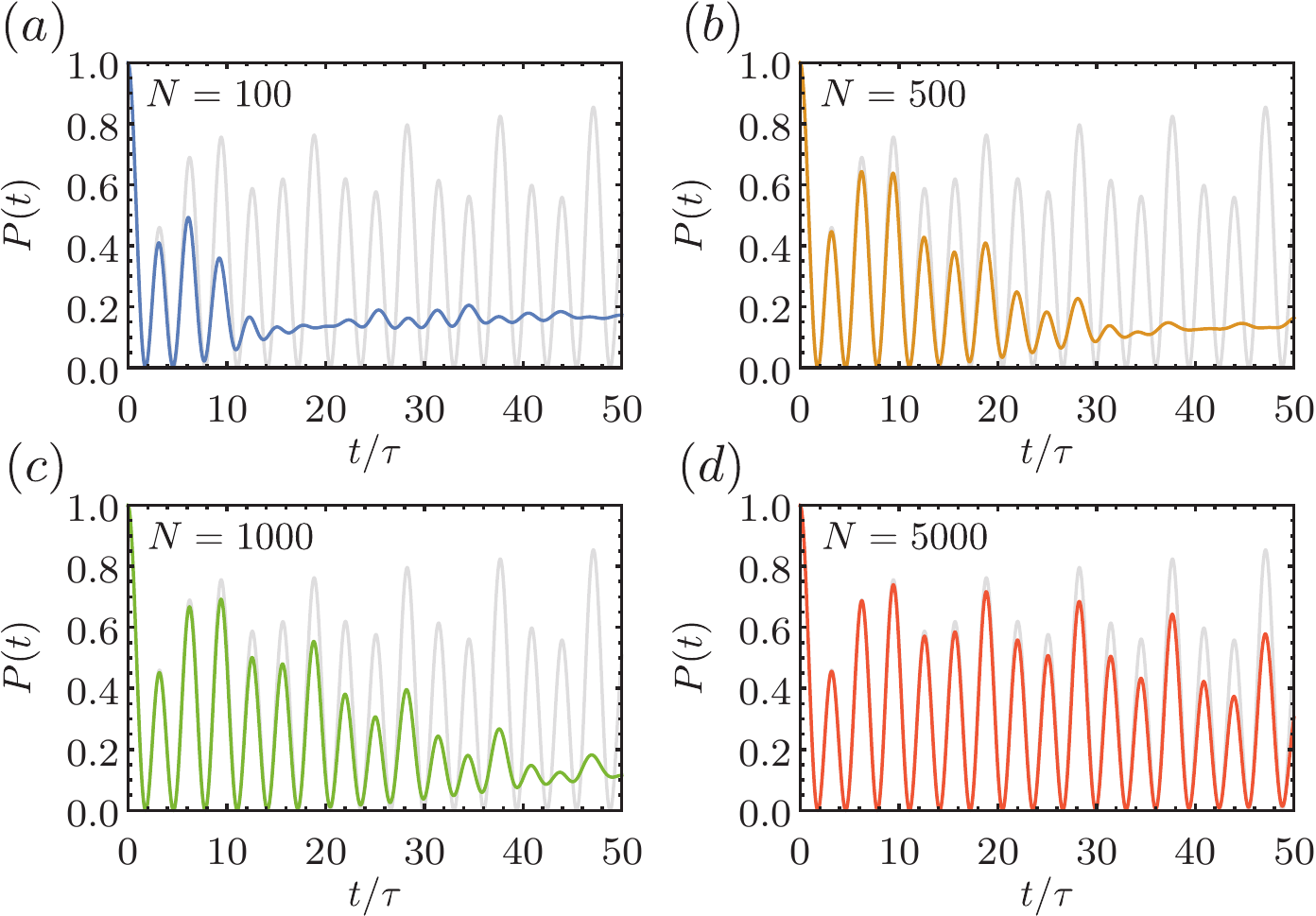}
\caption{Time evolution of the state $\ket{W}$ under the Hamiltonian $H_s$ for (a) $N = 100$, (b) $N = 500$, (c) $N = 1000$ and (d) $N = 5000$ particles after averaging over $10^5$ realizations with a Gaussian distribution and $k \sigma = 100$. The gray curve indicates the universal dynamics given by Eq. (\ref{eq:universal_function}).}
\label{fig:PW_diffN}
\end{figure}

First, we study the {\it  coherent dynamics} of the state $\ket{W}$ under the Hamiltonian $H_s$ alone. This Hamilton gives rise to a coupling between  $\ket{W}$ 
and the dark states $|D_{n}\rangle$.  Therefore, the quantity of interest is the probability  $P(t)$ to stay in the superatom state $\ket{W}$ under the coherent time 
evolution  after averaging over many experimental realizations, i.e.,  $P(t) = \big \langle \vert \bra{W} e^{-i H_s \hbar t} \ket{W} \vert^2 \big \rangle_{\rs{dis}}$. 
This probability can be evaluated 
numerically using exact diagonalization and averaging over different disorder realizations, and it is shown in Fig.~\ref{fig:PW_diffN}. 
Remarkably,  the dynamics features  robust revivals on the characteristic time scale $\tau = \pi / N \gamma$, which 
only damp out on the slower time scale $\tau_\text{dp}=  \sqrt{N} \tau$. Therefore, for increasing particle numbers, the amount of observable coherent oscillations increases. 
Finally, $P(t)$ saturates at a finite value $\sim 1/6$ for long times $t \gg \tau_\text{dp}$.  Note that in Fig.~\ref{fig:PW_diffN}  we chose a Gaussian density distribution
$n(x) = N \exp(- x^2/\sigma^2)/\sqrt{\pi \sigma^2}$; however, the above observations are independent of the atomic density profile as long as the the atomic cloud
is smooth on distances comparable to the optical wave length $\lambda$.

In the following, we provide an analytical analysis of this universal dynamics for the superatom state  $\ket{W}$. 
It turns out to be convenient to split the Hamiltonian  $H_s = H_{F} + H_{ B}$ into two parts, where $H_{F}$ ($H_{B}$) describes the virtual 
exchange of forward (backward) propagating photons, respectively. The part describing interaction between the atoms due to forward 
propagating photons is given by
\begin{equation}
H_{F} = \frac{\hbar \gamma}{2i} \int dx dy \, \text{sign}(x-y) \, e^{i k (x-y)} S^+(x) S^-(y) \, , 
\end{equation}

and analogously for $H_{B}$.
These Hamiltonians are exactly solvable \cite{sup} and the spectrum takes the form $E_\alpha = \frac{\hbar \gamma}{2} \cot \left(\frac{\alpha \pi}{2N} \right)$ 
with $\alpha$ an odd integer and $-N \leq \alpha < N$. Furthermore, the eigenstates are
\begin{equation}
\ket{\alpha, F} = \frac{1}{\sqrt{N}} \int dx e^{ikx} S^+(x) \exp \left(- i \frac{\pi \alpha}{N} F(x) \right) \ket{G}
\end{equation}
with the operator $F(x) = \int_{-\infty}^x dz \, \hat{n}_g(z)$ counting the number of ground state atoms on the left of position $x$; similar for $\ket{\alpha, B}$.

For a large atom number $N \gg 1$, only states with  $|\alpha | \ll N$ have significant overlap with the superatom state with  $\ket{W} = - \sum_\alpha 2 / (\pi \alpha )\ket{\alpha,F}$
and the energies reduce to $E_\alpha = N \hbar \gamma / \pi \alpha$.  As a result, the probability to remain in the bright state given only the forward propagating part of the Hamiltonian % $P^+_W(t) = \vert \bra{W} e^{-i H_+ t / \hbar} \ket{W} \vert^2$, 
is given by $\left[ \chi(t / \tau) \right]^2$ with $ \tau = \pi / N \gamma$ and
\begin{equation}
\chi(s) = \frac{8}{\pi^2} \sum_{n = 0}^\infty \frac{1}{(1+2n)^2} \cos \left( \frac{s}{1+2n} \right) \, . 
\label{eq:universal_function}
\end{equation}
It is this universal function, that  $P(t)$ approaches for an increasing number of atoms (see Fig.~\ref{fig:PW_diffN}). In order
to understand this observation, there are two important points to notice: First, only those states $\ket{\alpha, F}$ with small 
values of $\vert \alpha \vert$ have a significant overlap with $\ket{W}$. In addition,
 these states dominate the fast dynamical behavior with the characteristic energy scale $E_1 = \hbar \gamma N/\pi$. It is therefore 
 sufficient to restrict the analysis to low values of $\vert \alpha \vert$. Second, the states $\ket{\alpha, \rm{F}}$ with low values of 
 $\vert \alpha \vert$ become exact eigenstates of the full Hamiltonian $H_s$ with energy $E_\alpha$ in the limit of a large particle 
 number $N \to \infty$ and a smooth atomic density distribution with $\sigma \gg \lambda$. Then,  the universal dynamics 
 $P(t) = [\chi(t/ \tau)]^2$ is the asymptotic dynamical behavior for large particle numbers. Note that  the  precise condition of low 
 values of $\alpha$ reduces to $\vert \alpha \vert < \sigma / \lambda$ as shown below.

In order to prove the statement that the states $\ket{\alpha, F/B}$ with low values of $\alpha$ become exact eigenstates
in the limit $N \to \infty$ and $\lambda / \sigma \to 0$, we analyze the wave function overlap between 
eigenstates of $H_{F}$ with the eigenstates of $H_{ B}$, i.e., $h_{\alpha \beta} = \bra{\beta,B}H_{\rm{B}}\ket{\alpha, F}/E_{0} =  \braket{\beta,B}{\alpha,F}/\beta$, 
and the matrix element $\delta_\alpha = \bra{\alpha, F} H_{\rs B} \ket{\alpha, F} / E_1$. These dimensionless parameters take the form
\begin{align}
h_{\alpha \beta} &= \frac{1}{\beta}  \int \frac{dx}{N} e^{2ikx} \bra{G} \hat{n}_g(x) e^{- i \frac{\pi ( \alpha + \beta)}{N}F(x)} \ket{G} \, , \label{eq:overlap}\\
\delta_\alpha  &=  \int \frac{dx dy}{2i N^2} \, \text{sign}(x-y) \,  e^{2 i k (x-y)} \\  &  \hspace{1.5cm} \times \bra{G} \hat{n}_g(x) \hat{n}_g(y)   
 e^{- i\frac{\pi \alpha}{N} (F(x) - F(y))} \ket{G}  .  \nonumber
\end{align}
In the limit $N \to \infty$, we can replace the atomic density operator by its averaged expectation value $n(x)$ as the fluctuations in the density vanish with $1/\sqrt{N}$.
Then, the overlaps in $h_{\alpha \beta}$  reduce to the Fourier transformation of a smoothly varying function. 
Therefore, the overlap between states with low numbers of $\alpha, \beta$, i.e., $2 \vert k \vert \gg \pi \vert \alpha + \beta \vert / \sigma$, 
vanishes for  $\lambda / \sigma \to 0$; here, $\sigma$ denotes the characteristic size of the atomic cloud in general.
For example,  it vanishes exponentially for a Gaussian density distribution, while for a stepwise atomic distribution it vanishes as 
$(\lambda / \sigma)^2$. On the other hand, overlaps with $|\beta |\gtrsim \sigma/\lambda$ are suppressed by the factor $1/\beta$ in Eq.~(\ref{eq:overlap}).
Similarly, the expression for $\delta_\alpha$ reduces to  $\delta_\alpha =  c_0 \frac{\lambda}{\sigma} + c_1 \alpha \frac{\lambda^2}{\sigma^2} +\mathcal{O} ((\lambda / \sigma)^3) $,
with dimensionless parameters $c_0$ and $c_1$  of order unity, which only depend on the atomic 
density distribution $n(x)$.  The first term is an irrelevant shift in energy, while the second correction again vanishes as $(\lambda / \sigma)^2$.
In conclusion, we have demonstrated that the eigenstates $\ket{\alpha, F}$ with energy $E_\alpha$ become exact eigenstates of the 
full Hamiltonian $H_s$ for $|\alpha| < \sigma/\lambda$ in the limit $N \to \infty$ and $\lambda / \sigma \to 0$.

Next, we analyze the leading correction due to a finite number of particles $N$ in the regime $\lambda \ll \sigma$. The main influence are 
deviations from the mean density distribution $n(x)$ due to the random distribution of the particles within each experimental realization. 
These fluctuations lead to fluctuations of $h_{\alpha \beta}$ and $\delta_\alpha$. We illustrate this behavior for the overlap 
$w_{\alpha} = \braket{\alpha,B}{\alpha,F}$. The important quantity is the variance of these fluctuations, i.e.,   
$\Delta w_{\alpha} = \sqrt{\langle \vert w_{\alpha} \vert^2 \rangle_{\rs{dis}} - \langle \vert w_{\alpha} \vert \rangle_{\rs{dis}}^2}$,
and its leading contribution takes the form $\Delta w_{\alpha}=1/\sqrt{N}$. This result is derived using the 
general relation
\begin{equation}
\langle \hat{n}_g(x) \hat{n}_g(y) \rangle_{\rs{dis}} = \frac{N - 1}{N} n(x) n(y) + n(x) \delta(x-y)
\end{equation}
valid for a thermal gas on  distances studied in the present setup. Furthermore, the full distribution function 
for $|w_{\alpha}|^2$ can be derived (See Supplemental Material \cite{sup}), which leads to an exponential distribution
with a mean value $1/N$.

\begin{figure}[h]
\includegraphics[scale=1]{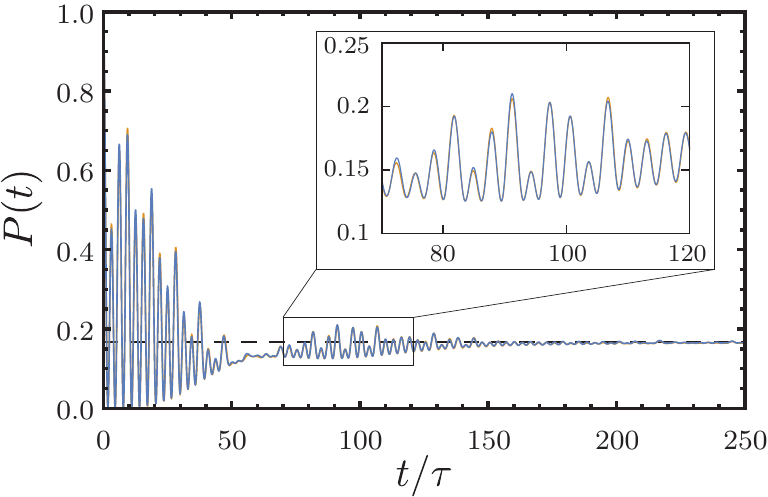}
\caption{Comparison between the numerically and analytically calculated time evolution of $\ket{W}$ under $H_s$. Blue (dashed) curve: Numerically calculated time evolution for $N = 1000$ particles averaged over $10^5$ realizations with a Gaussian distribution and $k \sigma = 100$. Orange (solid) curve: Analytical prediction for $\lambda/\sigma \ll 1$ and large $N$. (Inset:) Zoomed image of the dynamics for a better comparison between the numerically and analytically calculated time evolution. Note that there is hardly any visible difference.}
\label{fig:comparisonPW}
\end{figure}

The last step to understand the behavior of $P(t)$ is to derive the leading correction to the energies 
$E_{\alpha}$ using perturbation theory in the small parameter $w_{\alpha}$,
\begin{displaymath}
\frac{E^\pm_{\alpha}} { E_\alpha}=  1 \pm \vert w_{\alpha }\vert  \hspace{7pt} \mbox{with} \hspace{7pt}\ket{\alpha, \pm} =\frac{1}{\sqrt{2}} \left( \ket{\alpha,\rm{F}} \pm e^{i \phi_{\alpha}}  \ket{\alpha,\rm{B}}\right)
 \end{displaymath}
and $w_\alpha = \vert w_\alpha \vert e^{i \phi_\alpha}$. Therefore, the relevant energies of the Hamiltonian $H_{s}$ fluctuate within each experimental realization with a variance $\Delta E_{\alpha} = E_{\alpha}/\sqrt{N}$
giving rise to a characteristic dephasing rate $\tau_{\text{dp}} = \pi / \sqrt{N} \gamma$. This observation allows us to derive the leading dynamical behavior  $P(t)$ for the superatom states $\ket{W}$
by performing the average over many different experimental realizations using the knowledge on the distribution function of $|w_{\alpha}|^2$, 
\begin{widetext}
\begin{align}
P(t) &= \left\{\frac{8}{\pi^2} \sum_{n\geq0} \frac{2}{(2n+1)^2} \cos \left(\frac{t / \tau}{2n+1} \right) \left[1 - 2 f\left(\frac{t}{2 \tau_\text{dp}(2n+1)} \right) \right]\right\}^2 \nonumber \\
& \quad - \frac{16}{\pi^4} \sum_{n\geq 0} \left\lbrace \frac{2}{(2n+1)^4}  \left( \left[ 1- 2 f \left(\frac{t}{2 \tau_\text{dp}(2n+1)} \right)\right]^2 -  \left[1 - f \left(\frac{t}{\tau_\text{dp}(2n+1)} \right)\right]\right)\right\rbrace 
\label{eq:PWanalytical}
\end{align}
\end{widetext}
with $f(x) = x D(x)$ and the Dawson function $D(x) = e^{-x^2} \int_0^x dt \, e^{t^2}$ with the asymptotic limit $f(x \to \infty) = 1/2$. The first term in Eq.~(\ref{eq:PWanalytical}) is a modification of the universal function Eq.~(\ref{eq:universal_function}) which now includes damping on a time scale $\tau_\text{dp}$. For long times $t \gg \tau_\text{dp}$, the dynamics saturates at 
\begin{equation}
P(t) \xrightarrow[t \gg \tau_\text{dp}]{}\sum_{n=0}^\infty \left( \frac{2}{\pi (2n+1)} \right)^4 = \frac{1}{6}.
\end{equation}
In Fig.~(\ref{fig:comparisonPW}), we compare the numerically calculated $P(t)$ for $N = 1000$ averaged over $10^5$ realizations with a Gaussian density distribution with $k \sigma = 100$ and $P(t)$ given in Eq.~(\ref{eq:PWanalytical}), and find an excellent agreement.

Finally, we analyze the {\it dissipative dynamics}. 
The collective enhancement of the coupling between the forward propagating waveguide mode and the $\ket{W}$ state also implies an enhanced spontaneous emission rate
$\Gamma_{F} =  N \gamma $
into the forward direction. In turn, the spontaneous emission into the backward propagating mode, $\Gamma_{B} $, 
depends on the details of the atomic distribution within each experimental realization. In the case of a smooth atomic distribution whose characteristic length 
scale $\sigma$ is much larger than the optical wave length $\lambda$, the decay rate averaged over many realizations reduces to  $\Gamma_{B} \approx \gamma$,
and it accounts for the spontaneous emission of a single atom. 

It is important to point out that the characteristic time scale for the revivals in the coherent dynamics and the dissipative part are of the same order. On one hand, we conclude
that the coherent part always provides an intrinsic contribution to the dephasing of a superatom state. On the other hand,  it is important to disentangle the dissipative 
dynamics and the coherent part  for the experimental observation of the revivals. This goal can be achieved by quenching the spontaneous emission by tailoring the waveguide.

This approach is described in the following for an  {\it experimentally realistic } setup.  Such a setup exhibits in addition to the coupling to the waveguide 
naturally also a spontaneous emission into  free space with rate $\gamma_{0}$. As a first requirement, this decay must be comparable or smaller than the decay into the waveguide, i.e., $
\gamma \gtrsim \gamma_{0}$, which can be achieved in current experimental setups \cite{Goban2014,Vetsch2010}. Then, the coherent dynamics as well as the dephasing are collectively enhanced and 
appear on a time scale much faster than residual losses, $N\gamma\gg\sqrt{N} \gamma \gg \gamma_{0}$. Second, the initial preparation of the setup into the  superatom state $\ket{W}$ 
is achieved using a  
$\pi$ pulse with a time scale fast compared to the characteristic 
dynamics  $\tau$. As the Rabi frequency is also collectively enhanced, this condition reduces to  $\Omega \gg \gamma \sqrt{N}$, with the single-atom Rabi frequency $\Omega$. To bring out the effect of the coherent dynamics, we propose an experimental setup, where the atoms are coupled to a 
one-dimensional photonic crystal or Bragg grating such that the emission process is strongly suppressed due to the opening of a band gap while the virtual 
photons mediating the exchange interaction can still propagate outside the photonic band gap.  In order to satisfy this condition, the size of the photonic band gap  $\Delta$
is required to be in the range  $\Gamma_{F} \ll \Delta \ll 2\pi c  / \sigma$, where $\sigma$ is the characteristic size of the system. The lower bound results from the fact 
that the emitted photon has a Lorentzian spectrum. 
The upper bound derives from the condition that the virtual photons should be able to propagate with a linear dispersion, such that the initial form of the exchange 
Hamiltonian is unaffected. For typical quantum optics experiments,
the system size is in the micrometer regime which relates to a mode 
spacing of the virtual photons of a few THz. In addition, the  decay into the waveguide is typically in the range of MHz, providing the enhanced decay rate $\Gamma_{F}=N\gamma$  in the lower GHz regime for $N \sim 10^4$ atoms.  This requires the width of the gap to be of the size of a few ten to hundred GHz. Such gratings have been
 produced, for example, in germanosilicate optical fibers \cite{Meltz1989} and they have been specifically designed for quantum optics experiments \cite{Yu2014}.

In conclusion, we demonstrated that the nonequilibrium dynamics of a quantum many-body system, consisting of atoms coupled to a 
one-dimensional waveguide, can exhibit a highly nontrivial universal dynamics characterized by revivals and eventually a saturation 
on $1/6$. This observation is independent on the averaged atomic distribution $n(x)$, and it
becomes more pronounced for increasing particle numbers. In the present analysis, we chose a fixed number of atoms within each experimental
realization. However, it is straightforward to derive that for a Poisson distributed number of atoms, the only modification in the
dynamics is the enhancement of the dephasing rate by a factor $\sqrt{2}$, while the revivals and the saturation remain unaffected. 
We expect that similar phenomena can also appear in free space for two-level systems strongly coupled to a single highly focused light mode.

 {\bf Acknowledgments}
We thank Christoph Tresp for helpful discussions and comments on the manuscript. We thank Christoph Braun and Asaf Paris-Mandoki for discussions at early stages of this work. This work is supported by the European Union under the ERC consolidator grants SIRPOL (Grant No. 681208) and RYD-QNLO (Grant No. 771417), and by the Deutsche Forschungsgemeinschaft (DFG) within the research unit FOR 2247 and the SPP 1929 GiRyd (Project No. HO
4787/3-1).

\bibliographystyle{apsrev4-1}
%\bibliography{Refs}



\section*{Supplementary material}

# Supplemental Material: Emergent universal dynamics for an atomic cloud coupled to an optical wave-guide

Jan Kumlin,[1, *] Sebastian Hofferberth,[2] and Hans Peter Büchler[1]

[1] *Institute for Theoretical Physics III, University of Stuttgart, 70550 Stuttgart, Germany*
[2] *Department of Physics, Chemistry and Pharmacy,
University of Southern Denmark, 5320 Odense, Denmark*
(Dated: March 23, 2018)

## DERIVATION OF MASTER EQUATION

Here, we give a derivation of the full master equation (5) in the main text. The derivation closely follows the methods and approximations used in [S1, S2] with minor modifications for our setup and is repeated here for the sake of completeness. We start with the Hamiltonian of the full system in the dipole and rotating-wave approximation given by

$$H = \int \frac{dq}{2\pi} \hbar\omega_q a_q^\dagger a_q + \hbar\omega_0 \int dx\, \psi_e^\dagger(x)\psi_e(x) - \hbar\sqrt{\gamma} \int dx\, \left[\mathcal{E}^\dagger(x)S^-(x) + S^+(x)\mathcal{E}(x)\right]. \tag{S1}$$

The bosonic operators $a_q^\dagger$ account for the creation of a wave guide mode with momentum $q$ and $\omega_q$ is the dispersion relation, which we will assume to be linear, $\omega_q = c|q|$. The second term simply gives the total energy of the atomic system. Finally, the last term describes the interaction, with an effective mode coupling $\gamma$, between the atoms and the photonic modes where $S^+(x) = \psi_e^\dagger(x)\psi_g(x)$ and the electric field operator is given by

$$\mathcal{E}^\dagger(x) = -i\sqrt{c} \int \frac{dq}{2\pi} a_q^\dagger e^{-iqx}. \tag{S2}$$

In a first step, we go over to the interaction picture with the interaction Hamiltonian

$$H_I = i\hbar\sqrt{\gamma c} \int dx \int \frac{dk}{2\pi} \left(a_k^\dagger S^-(x) e^{-ikx} e^{i(\omega_k-\omega_0)t} - S^+(x) a_k e^{ikx} e^{-i(\omega_k-\omega_0)t}\right). \tag{S3}$$

The equation of motion for the annihilation operator of some mode $q$ is given by

$$\dot{a}_q(t) = -\frac{i}{\hbar}[a_q(t), H_I] = \sqrt{\gamma c} \int dx\, S^-(x,t) e^{-iqx} e^{i(\omega_q-\omega_0)t}. \tag{S4}$$

and has the solution

$$a_q(t) = a_q(0) + \sqrt{\gamma c} \int dx \int_0^t ds\, S^-(x,s) e^{i(\omega_q-\omega_0)s} e^{-iqx}. \tag{S5}$$

Accordingly, the equation of motion of some operator $O$ which acts only on the atomic degrees of freedom reads

$$\dot{O}(t) = \sqrt{\gamma c} \int \frac{dq}{2\pi} \int dx\, \left(a_q^\dagger(t) e^{i(\omega_q-\omega_0)t} e^{-iqx}[O(t), S^-(x,t)] - e^{-i(\omega_q-\omega_0)t} e^{iqx}[O(t), S^+(x,t)] a_q(t)\right). \tag{S6}$$

Plugging in eq. (S5), we arrive at

$$\begin{aligned}\dot{O}(t) = &\sqrt{\gamma c} \int dx \int \frac{dq}{2\pi} \left(a_q^\dagger(0)[O(t), S^-(x,t)] e^{i(\omega_q-\omega_0)t} e^{-iqx} - [O, S^+(x,t)] a_q(0) e^{-i(\omega_q-\omega_0)t} e^{iqx}\right) \\ &+ \gamma c \int dx \int dy \int \frac{dq}{2\pi} \int_0^t ds\, \Big(S^+(y,s)[O(t), S^-(x,t)] e^{i(\omega_q-\omega_0)(t-s)-iq(x-y)} \\ &\qquad\qquad -[O(t), S^+(x,t)] S^-(y,s) e^{-i(\omega_q-\omega_0)(t-s)+iq(x-y)}\Big). \end{aligned}\tag{S7}$$

Before we further manipulate these terms, we use the assumption that the dispersion relation is linear and replace $c \int dq = \sum_{\lambda=\pm} \int_{\omega_0-\theta}^{\omega_0+\theta} d\omega$, where $\lambda$ denotes the direction of propagation of the mode with frequency $\omega$ and the integration is restricted to a narrow frequency band with cutoff $\theta \ll \omega_0$ around the resonance frequency $\omega_0$. This

restriction is required by the use of the rotating-wave approximation. The Markov approximation then consists of taking the limit $\theta \to \infty$ assuming that the correlation times of the photonic operators are much smaller than the typical time scale on which atomic operators change. In our setup, this approximation is valid for $\gamma \ll \theta \ll \omega_0$. It is then possible to simplify the last two lines of eq. (S7) using

$$\int_0^t ds \int \frac{d\omega}{2\pi} e^{i(\omega-\omega_0)(t-s-\lambda(x-y)/c)} e^{-i\lambda\omega_0(x-y)/c} S^+(y,s) = \int_0^t ds \int_{-\theta}^{\theta} \frac{d\nu}{2\pi} e^{i\nu(t-s-\lambda(x-y)/c)} e^{-i\lambda\omega_0(x0y)/c} S^+(y,s)$$

$$= \int_0^t ds\, \delta(t-s-\lambda(x-y)/c) e^{-i\lambda\omega_0(x-y)/c} S^+(y,s)$$

$$= \begin{cases} \theta(\lambda(x-y)) e^{-i\lambda\omega_0(x-y)/c} S^+(y, t-\lambda(x-y)/c) &, x \neq y \\ \frac{1}{2} S^+(y,t) &, x = y \end{cases} \quad (S8)$$

for $ct > |x-y|$ and with the Heaviside function $\theta(x) = 1$ for $x > 0$ and $\theta(x) = 0$ for $x \leq 0$. Further, we neglect any retardation effects, thereby replacing $S^+(y, t - \lambda(x-y)/c) \approx S^+(y,t)$ which is valid for $|x-y|_{\max} \ll c/\gamma$. We can then rewrite eq. (S7) as

$$\dot{O}(t) = \sqrt{\gamma c} \int dx \int \frac{dq}{2\pi} \left( a_q^\dagger(0)[O(t), S^-(x,t)] e^{i(\omega_q-\omega_0)t} e^{-iqx} - [O, S^+(x,t)] a_q(0) e^{-i(\omega_q-\omega_0)t} e^{iqx} \right)$$

$$+ \frac{\gamma}{2} \int dx \left( S^+(x,t)[O(t), S^-(x,t)] - [O(t), S^+(x,t)] S^-(x,t) \right)$$

$$+ \gamma \int dx \int dy \sum_{\lambda=\pm} \theta(\lambda(x-y)) \left( S^+(y,t)[O(t), S^-(x,t)] e^{-i\lambda\omega_0(x-y)/c} - [O(t), S^+(x,t)] S^-(y,t) e^{i\lambda\omega_0(x-y)/c} \right) \quad (S9)$$

In the next step, we aim at deriving a master equation for the time evolution of the density matrix $\rho$ of the atomic system alone. In order to do so, we note that the expectation value of the operator $O$ is given by $\langle O(t) \rangle = \text{Tr}(R(0)O(t))$ where $R(0)$ is the initial density matrix of the whole system and the trace includes both the atomic and photonic degrees of freedom. Initially, we assume the system to be in the state $R(0) = |0\rangle\langle 0| \otimes \rho(0)$, where $|0\rangle$ denotes the photonic vacuum and $\rho(0)$ is the initial density matrix of the atomic subsystem. Further, we can also write $\langle O(t) \rangle = \text{Tr}(R(t)O) = \text{Tr}_S(\rho(t)O)$ where the trace in the second expression only runs over the atomic degrees of freedom. Then, it is possible to derive a master equation describing the time evolution of the atomic density matrix $\rho$ which reads

$$\dot{\rho}(t) = \gamma \int dx \left( S^-(x)\rho(t)S^+(x) - \frac{1}{2}\{S^+(x)S^-(x), \rho(t)\} \right)$$

$$+ \gamma \int dx \int dy \sum_{\lambda=\pm} \theta(\lambda(x-y)) \left( e^{-i\lambda\omega_0(x-y)/c} (S^-(x)\rho(t)S^+(y) - \rho(t)S^+(y)S^-(x)) \right.$$

$$\left. + e^{i\lambda\omega_0(x-y)/c} (S^-(y)\rho(t)S^+(x) - S^+(x)S^-(y)\rho(t)) \right) \quad (S10)$$

This can now be rewritten as

$$\dot{\rho}(t) = -i\gamma \int dx \int dy \, \sin(\omega_0|x-y|/c) \left[ S^+(x)S^-(y), \rho(t) \right]$$

$$+ 2\gamma \int dx \int dy \, \cos(\omega_0(x-y)/c) \left( S^-(x)\rho(t)S^+(y) - \frac{1}{2}\{S^+(x)S^-(y), \rho(t)\} \right) \quad (S11)$$

Defining the wave vector $k = \omega_0/c$ and the Hamiltonian $H_s = \hbar\gamma \int dx \int dy \, \sin(k|x-y|) S^+(x)S^-(y)$, we arrive at eq. (5) in the main text.

## SPECTRUM AND EIGENSTATES

In this section, we show that the eigenenergies of $H_F = \frac{\hbar\gamma}{2i} \int dx\, dy \, \text{sign}(x-y) e^{ik(x-y)} S^+(x)S^-(y)$ are given by

$$E_\alpha = \frac{\hbar\gamma}{2} \cot\left(\frac{\pi\alpha}{2N}\right) \quad (S12)$$

with an odd integer $-N \leq \alpha < N$. The eigenstates are given by

$$|\alpha, \text{F}\rangle = \frac{1}{\sqrt{N}} \int dx \, e^{ikx} S^+(x) \exp\left(-i\frac{\pi\alpha}{N} F(x)\right) |G\rangle \qquad (\text{S13})$$

with $F(x) = \int_{-\infty}^{x} dz \, \hat{n}_g(z)$.

In order to do so, we apply the state (S13) onto the Hamiltonian $H_+$ and recall the definition of the state $|G\rangle = 1/\sqrt{N!} \prod_i \psi_g^\dagger(x_i) |0\rangle$ and the operators $S^+(x) = \psi_e^\dagger(x)\psi_g(x)$ and $S^-(x) = \psi_g^\dagger(x)\psi_e(x)$. We then make an ansatz for the eigenstate of the form

$$|\alpha, \text{F}\rangle = \frac{1}{\sqrt{N}} \sum_{n=1}^{N} e^{ikx_n} S^+(x_n) e^{-i\frac{\pi\alpha n}{N}} |G\rangle, \qquad (\text{S14})$$

and rewrite the Hamiltonian as

$$H_\text{F} = \frac{\hbar\gamma}{2i} \sum_{n=1}^{N} \sum_{l=1}^{N} \text{sign}(x_n - x_l) e^{ik(x_n - x_l)} S^+(x_n) S^-(x_l). \qquad (\text{S15})$$

Note that in the following, we always assume the positions of the atoms to be ordered, that is $x_1 \leq x_2 \leq \ldots \leq x_N$. Using that $S^-(x_n) S^+(x_l) |G\rangle = \delta_{nl} |G\rangle$, we get

$$H_\text{F} |\alpha, \text{F}\rangle = \frac{\hbar\gamma}{2i\sqrt{N}} \sum_{n,l,m=1}^{N} e^{ikx_m} e^{-i\frac{\pi\alpha m}{N}} \text{sign}(x_n - x_l) e^{ik(x_n - x_l)} S^+(x_n) S^-(x_l) S^+(x_m) |G\rangle$$

$$= \frac{\hbar\gamma}{2i\sqrt{N}} \sum_{n,l=1}^{N} e^{ikx_n} e^{-i\frac{\pi\alpha l}{N}} \text{sign}(x_n - x_l) S^+(x_n) |G\rangle$$

$$= \frac{\hbar\gamma}{2i\sqrt{N}} \sum_{n=1}^{N} e^{ikx_n} \left( \sum_{l=1}^{n-1} e^{-i\frac{\pi\alpha l}{N}} - \sum_{l=n+1}^{N} e^{-i\frac{\pi\alpha l}{N}} \right) S^+(x_n) |G\rangle$$

$$= \frac{\hbar\gamma}{2i\sqrt{N}} \sum_{n=1}^{N} e^{ikx_n} S^+(x_n) |G\rangle \frac{e^{-i\frac{\pi\alpha n}{N}}(1 + e^{-i\frac{\pi\alpha}{N}}) - e^{-i\frac{\pi\alpha}{N}}(1 + e^{-i\pi\alpha})}{e^{-i\frac{\pi\alpha}{N}} - 1}. \qquad (\text{S16})$$

From the last line, it is clear that if $\alpha$ is an odd integer, the second term in the numerator vanishes and $|\alpha, \text{F}\rangle$ is indeed an eigenstate of $H_\text{F}$. The corresponding eigenvalue is then

$$E_\alpha = \frac{\hbar\gamma}{2i} \frac{e^{-i\frac{\pi\alpha}{N}} + 1}{e^{-i\frac{\pi\alpha}{N}} - 1} = \frac{\hbar\gamma}{2} \cot\left(\frac{\pi\alpha}{2N}\right). \qquad (\text{S17})$$

From (S17) it is also clear that $\alpha$ is limited to $-N \leq \alpha < N$ due to the periodicity of the eigenvalues.

For the left-moving Hamiltonian $H_\text{B} = -\frac{1}{2i} \sum_{n,l} \text{sign}(x_n - x_l) e^{-ik(x_n - x_l)} S^+(x_n) S^-(x_l)$, we make the ansatz

$$|\alpha, \text{B}\rangle = \frac{1}{\sqrt{N}} \sum_{n=1}^{N} e^{-ikx_n} S^+(x_n) e^{i\frac{\pi\alpha n}{N}} |G\rangle. \qquad (\text{S18})$$

The proof that this state is an eigenstate of $H_\text{B}$ follows along the same lines as above and the eigenvalue is also given by $E_\alpha = \hbar\gamma/2 \cot(\pi\alpha/2N)$.

## UNIVERSAL DYNAMICS

In order to derive the universal dynamics given by (9) in the main text, we consider the Hamiltonian $H_F$ with its eigenstates $|\alpha, \text{F}\rangle$ and -energies $E_\alpha$ only. The overlap with $|W\rangle$ is given by

$$\langle W | \alpha, \text{F} \rangle = \frac{1}{N} \sum_{j=1}^{N} e^{-i\alpha j/N} = -\frac{2}{N} \frac{e^{-i\alpha}}{e^{i\alpha/N} - 1} \approx -\frac{2}{\pi\alpha}. \qquad (\text{S19})$$

The time evolution $P(t)$ in the limit $N \to \infty$ can then be calculated as

$$P(t) = \left| \sum_\alpha e^{-iE_\alpha} |\langle W|\alpha, \mathrm{F}\rangle|^2 \right|^2$$

$$\approx \left| \sum_\alpha e^{-i\frac{\gamma Nt}{\pi\alpha}} \left(\frac{2}{\pi\alpha}\right)^2 |\langle W|\alpha, \mathrm{F}\rangle|^2 \right|^2$$

$$= \left| \sum_{j=0}^\infty \left(\frac{2}{\pi(2j+1)}\right)^2 \cos\left(\frac{\gamma Nt}{\pi(2j+1)}\right) \right|^2 \quad \text{(S20)}$$

**PERTURBATION THEORY**

As mentioned in the text, we define two dimensionless variables

$$h_{\alpha\beta} = \frac{\langle \beta, \mathrm{B}|\alpha, \mathrm{F}\rangle}{\beta} = \frac{1}{N\eta} \int dx\, e^{2ikx} \langle G|\, \hat{n}_g(x) e^{-i\frac{\pi(\alpha+\beta)}{N} F(x)} |G\rangle \quad \text{(S21)}$$

$$\delta_\alpha = \frac{\langle \alpha, \mathrm{F}|\, H_\mathrm{B}\, |\alpha, \mathrm{F}\rangle}{E_1} = -\frac{1}{2iN^2} \int dxdy\, e^{2ik(x-y)} \mathrm{sign}(x-y) \langle G|\, \hat{n}_g(x)\hat{n}_g(y) e^{-i\frac{2\pi\alpha}{N}(F(x)-F(y))} |G\rangle \,. \quad \text{(S22)}$$

Since the positions of the atoms are randomly distributed in each experimental realization, the ground state density distribution $\hat{n}_g(x)$ and consequently the variables $h_{\alpha\beta}$ and $\delta_\alpha$ will fluctuate. In the following, we will calculate the leading order behavior of the mean values and the fluctuations of these variables in the limits $k\sigma \to \infty$ and $N \to \infty$. To this end, we split the density distribution as $\hat{n}_g(x) = n(x) + \delta\hat{n}_g(x)$ where $n(x) = \langle \hat{n}_g(x)\rangle_\mathrm{dis}$ is the mean density distribution and thus $\langle \delta\hat{n}_g(x)\rangle_\mathrm{dis} = 0$. Further, we will make use of

$$\langle \hat{n}_g(x)\hat{n}_g(y)\rangle_\mathrm{dis} = g^{(2)}(x,y)n(x)n(y) + \delta(x-y)n(x) \quad \text{(S23)}$$

with the two-body correlation function $g^{(2)}(x,y)$. Since the atoms are uncorrelated on distances comparable to the optical wave length, we have $g^{(2)} = 1 - 1/N$ for a fixed atom number $N$. Similarly, we get

$$\langle \delta\hat{n}_g(x)\delta\hat{n}_g(y)\rangle_\mathrm{dis} = (g^{(2)}(x,y) - 1)n(x)n(y) + \delta(x-y)n(x) = -\frac{n(x)n(y)}{N} + \delta(x-y)n(x)\,. \quad \text{(S24)}$$

**Mean values**

We start with calculating the mean value of $h_{\alpha\beta}$. Therefore, we note that we can write

$$\overline{h_{\alpha\beta}} = \frac{1}{N\beta} \int dx\, e^{2ikx} \left\langle \hat{n}_g(x) e^{-i\frac{\pi(\alpha+\beta)}{N}\int_{-\infty}^x dz\, \hat{n}_g(z)} \right\rangle_\mathrm{dis} = \frac{1}{N\beta} \int dx\, e^{2ikx} \left(\frac{iN}{\pi(\alpha+\beta)}\right) \partial_x \left\langle e^{-i\frac{\pi(\alpha+\beta)}{N}\int_{-\infty}^x dz\, \hat{n}_g(z)} \right\rangle_\mathrm{dis} \quad \text{(S25)}$$

where we assume that $\hat{n}_g(-\infty) = 0$. Using $\hat{n}_g(x) = n(x) + \delta\hat{n}_g(x)$, we can rewrite

$$\left\langle e^{-i\frac{\pi(\alpha+\beta)}{N}\int_{-\infty}^x dz\, \hat{n}_g(z)} \right\rangle_\mathrm{dis} = e^{-i\frac{\pi(\alpha+\beta)}{N}N(x)} \sum_{m=0}^\infty \frac{(i\pi(\alpha+\beta))^m}{N^m m!} \int dz_1 \cdots dz_m \theta(x-z_1) \cdots \theta(x-z_m) \langle \delta\hat{n}_g(z_1) \cdots \delta\hat{n}_g(z_m)\rangle_\mathrm{dis}, \quad \text{(S26)}$$

where we defined $N(x) = \int_{-\infty}^x dz\, n(z)$ and $\theta(x)$ is the Heaviside function. The correlator can now be evaluated using Isserlis' theorem (in quantum field theory also known as Wick's theorem) which states that each $m$-point correlator of a variable with zero mean can be written as a sum over a product of all possible combinations of two-point correlators for even $m$. Since the total number of terms in the sum is given by $(m-1)!! = (2n-1)!! = (2n)!/(2^n n!)$, we can write

$$\sum_{m=0}^\infty \frac{(i\pi(\alpha+\beta))^{2n}}{N^{2n} 2^n n!} \left(\int dz_1 dz_2\, \theta(x-z_1)\theta(x-z_2)\langle \delta\hat{n}_g(z_1)\delta\hat{n}_g(z_2)\rangle\right)^n = e^{-\frac{1}{2}\left(\frac{\pi(\alpha+\beta)}{N}\right)^2 \left(N(x) - \frac{N(x)^2}{N}\right)} \quad \text{(S27)}$$

Such that finally we have

$$\overline{h_{\alpha\beta}} = \frac{1}{N\beta} \int dx \, e^{2ikx} \left( \frac{iN}{\pi(\alpha+\beta)} \right) \partial_x e^{-i\frac{\pi(\alpha+\beta)}{N}N(x) - \frac{1}{2}\left(\frac{\pi(\alpha+\beta)}{N}\right)^2 \left(N(x) - \frac{N(x)^2}{N}\right)}. \tag{S28}$$

Performing the derivative, we finally get

$$\overline{h_{\alpha\beta}} = \frac{1}{N\beta} \int dx \, e^{2ikx} n(x) \left[ 1 - \frac{i\pi(\alpha+\beta)}{2N}\left(1 - \frac{2N(x)}{N}\right) \right] e^{-i\frac{\pi(\alpha+\beta)}{N}N(x) - \frac{1}{2}\left(\frac{\pi(\alpha+\beta)}{N}\right)^2 \left(N(x) - \frac{N(x)^2}{N}\right)}. \tag{S29}$$

This expression is now simply the Fourier transform of an (in principle very complicated) function. If however $n(x)$ and $N(x)$ are smooth functions and $2k\sigma \gg \pi|\alpha+\beta|$, the mean value $\overline{h_{\alpha\beta}}$ is strongly suppressed. As an example, we consider a Gaussian distribution of the particles with $n(x) = N/\sqrt{2\pi\sigma^2} e^{-x^2/2\sigma^2}$ and $N(x) = N/2(1 + \text{erf}(x/\sqrt{2}\sigma))$, where $\text{erf}(x)$ is the error function. Focusing on the limit $N \to \infty$, where we can safely neglect terms of order $1/N$, we get

$$\overline{h_{\alpha\beta}} = \frac{1}{\beta} \int dx \, e^{2ikx} \frac{e^{-x^2/2\sigma^2}}{\sqrt{2\pi\sigma^2}} e^{-i\frac{\pi(\alpha+\beta)}{2}(1+\text{erf}(x/\sqrt{2}\sigma))}, \tag{S30}$$

which vanishes exponentially in the limit $2k\sigma \gg \pi|\alpha+\beta|$. On the other hand, for $|\beta| \geqslant \sigma/\lambda$, this value is suppressed by the factor $1/\beta$.

For the mean value of $\delta_\alpha$, we write

$$\overline{\delta_\alpha} = -\frac{1}{2iN^2} \int dxdy \, e^{2ik(x-y)} \text{sign}(x-y) \frac{N^2}{(2\pi\alpha)^2} \partial_x \partial_y \left\langle e^{-i\frac{2\pi\alpha}{N}\int dz \hat{n}_g(z)(\theta(x-z)-\theta(y-z))} \right\rangle_{\text{dis}}$$

$$= -\frac{1}{2iN^2} \int dxdy \, e^{2ik(x-y)} \text{sign}(x-y) \frac{N^2}{(2\pi\alpha)^2} \times$$

$$\times \partial_x \partial_y e^{-i\frac{2\pi\alpha}{N}(N(x)-N(y))} e^{-\frac{1}{2}\left(\frac{2\pi\alpha}{N}\right)^2 \int dz_1 dz_2(\theta(x-z_1)-\theta(y-z_1))(\theta(x-z_2)-\theta(y-z_2))\langle\delta\hat{n}_g(z_1)\delta\hat{n}_g(z_2)\rangle}$$

$$= -\frac{1}{2iN^2} \int dxdy \, e^{2ik(x-y)} \text{sign}(x-y) \frac{N^2}{(2\pi\alpha)^2} \partial_x \partial_y e^{-i\frac{2\pi\alpha}{N}(N(x)-N(y))} e^{-\frac{1}{2}\left(\frac{2\pi\alpha}{N}\right)^2 \left(\int dz(\theta(x-z)-\theta(y-z))^2 n(z) - \frac{1}{N}(N(x)-N(y))^2\right)}. \tag{S31}$$

Since we are again only interested in the limit where $N \to \infty$, the expression reduces to

$$\overline{\delta_\alpha} \approx -\frac{1}{2i} \int dxdy \, e^{2ik(x-y)} \text{sign}(x-y) \frac{n(x)n(y)}{N^2} e^{-i\frac{2\pi\alpha}{N}(N(x)-N(y))}$$

$$= -\int_{-\infty}^{\infty} dx \int_{-\infty}^{x} dy \sin(2k(x-y) - 2\pi\alpha(N(x)-N(y))/N) \frac{n(x)n(y)}{N^2}, \tag{S32}$$

where we assumed that $n(-x) = n(x)$ and consequently $N(-x) = N - N(x)$. Given a characteristic length $\sigma$ of the atomic distribution and the condition that $k\sigma \gg \pi|\alpha|$, we may neglect the second term in phase of the sine and write

$$\overline{\delta_\alpha} \approx -\int_{-\infty}^{\infty} dx \int_{-\infty}^{x} dy \sin(2k(x-y))) \frac{n(x)n(y)}{N^2}. \tag{S33}$$

In the following, we will estimate the value of this integral and derive the behavior in leading order for $k\sigma \gg 1$. Since we are only interested in realistic density distributions, we may estimate the integral by an exponential distribution $f(x) \sim e^{-x^2/2\sigma^2}$ which we make broad and large enough such that $f(x) > n(x)$ for all $x$. Note that the width of the exponential distribution is assumed to be some multiple of the width of the original distribution. Then, the integral can be calculated analytically as

$$\int_{-\infty}^{\infty} dx \int_{-\infty}^{x} dy \sin(2k(x-y))) \frac{n(x)n(y)}{N^2} < \int_{-\infty}^{\infty} dx \int_{-\infty}^{x} dy \sin(2k(x-y))) \frac{e^{-(x^2+y^2)/2\sigma^2}}{2\pi\sigma^2}$$

$$= \int_{-\infty}^{\infty} dx \int_{-\infty}^{\infty} dy \, \theta(x-y) \sin(2k(x-y)) \frac{e^{-(x^2+y^2)/2\sigma^2}}{2\pi\sigma^2}$$

$$= \int dR \frac{e^{-R^2/2}}{\sqrt{2\pi}} \int dr \, \theta(r) \sin(\sqrt{2}k\sigma r) \frac{e^{-r^2/2}}{\sqrt{2\pi}}$$

$$= \sqrt{2}D(k\sigma) \sim \frac{1}{\sqrt{2}\,k\sigma}, \tag{S34}$$

where $D(x) = e^{-x^2} \int_0^x dt\, e^{t^2}$ is the Dawson function with the asymptotic expansion

$$D(x) = \sum_{n=0}^{\infty} \frac{(2n-1)!!}{2^{n+1} x^{2n+1}} = \frac{1}{2x} + \mathcal{O}(x^{-3}). \tag{S35}$$

Thus, the leading order behavior of $\overline{\delta_\alpha}$ in the limit $N \to \infty$ and $k\sigma \gg \pi|\alpha|$ is bounded by

$$\overline{\delta_\alpha} \sim -\frac{1}{\sqrt{2}\, k\sigma}. \tag{S36}$$

and consequently vanishes in the limit $k\sigma \gg \pi|\alpha|$.

### Fluctuations

*Distribution of $|w|^2$*

Here, we prove that in the limit $k\sigma \gg \pi|\alpha|$, $X = |w|^2 = |w_\alpha|^2$ follows an exponential distribution $p(X) = \lambda e^{-\lambda X}$ with mean $\bar{X} = 1/\lambda$. The exponential distribution can be characterized by its moments $\bar{X}^n = n!/\lambda^n$. The $n$-th moment of $|w|^2$ is given by

$$\overline{|w|^{2n}} = \frac{1}{N^{2n}} \int dx_1 \cdots dx_{2n}\, e^{-2ik(x_1-x_2)} \cdots e^{-2ik(x_{2n-1}-x_{2n})} \left\langle \hat{n}_g(x_1) \cdots \hat{n}_g(x_{2n}) e^{-i\frac{\pi\alpha}{N}(F(x_1)-F(x_2)+\cdots+F(x_{2n-1})-F(x_{2n}))} \right\rangle_{\text{dis}}$$

$$= \frac{1}{N^{2n}} \int dx_1 \cdots dx_{2n}\, e^{-2ik(x_1-x_2)} \cdots e^{-2ik(x_{2n-1}-x_{2n})} \sum_{m=0}^{\infty} \frac{(-i\pi\alpha)^m}{N^m m!} \int dz_1 \cdots dz_m h(\mathbf{x}, z_1) \cdots h(\mathbf{x}, z_m) \times$$

$$\times \langle \hat{n}_g(x_1) \cdots \hat{n}_g(x_{2n}) \hat{n}_g(z_1) \cdots \hat{n}_g(z_m) \rangle_{\text{dis}}, \tag{S37}$$

where we defined $h(\mathbf{x}, z) = \theta(x_1 - z) - \theta(x_2 - z) + \cdots + \theta(x_{2n-1} - z) - \theta(x_{2n} - z)$. In the next step, we express the density operator using $\hat{n}_g(x) = n(x) + \delta \hat{n}_g(x)$ in order to apply Isserlis' theorem. Further, we point out that whenever we pair operators with variables $x$ and $z$, some phase factor $e^{ikx}$ remains and this term is strongly suppressed in the limit $k\sigma \gg \pi\alpha$. Thus, the only contribution with no phase factors comes from the terms where only $x$ variables are paired, that is

$$\overline{|w|^{2n}} = \frac{1}{N^{2n}} \int dx_1 \cdots dx_{2n}\, e^{-2ik(x_1-x_2)} \cdots e^{-2ik(x_{2n-1}-x_{2n})} e^{-i\frac{\pi\alpha}{N}(N(x-1)-N(x_2)+\cdots N(x_{2n-1})-N(x_{2n}))} \langle \delta\hat{n}_g(x_1) \cdots \delta\hat{n}_g(x_{2n}) \rangle_{\text{dis}}. \tag{S38}$$

In this expression, only those terms where odd indices are paired with even ones do not have a phase factor. For $2n$ indices, there are in total $n!$ possibilities for this kind of pairing and we are left with

$$\overline{|w|^{2n}} = \frac{n!}{N^{2n}} \int dx_1 \cdots dx_{2n}\, n(x_1) \cdots n(x_{2n}) = \frac{n!}{N^{2n}} \left( \int dx\, n(x) \right)^n = \frac{n!}{N^n}, \tag{S39}$$

which shows that indeed $|w|^2$ follows an exponential distribution with mean $\overline{|w|^2} = 1/N$. Consequently, the variance of the complex variable $w_\alpha$ is given by

$$(\Delta w_\alpha)^2 = \langle |w_\alpha|^2 \rangle_{\text{dis}} - \langle |w_\alpha| \rangle_{\text{dis}}^2 = \frac{1}{N}. \tag{S40}$$

*Fluctuations in $\delta_\alpha$*

In a next step, we calculate the fluctuations in the variable $\delta_\alpha$. Since this variable is real, the variance is given by

$$(\Delta \delta_\alpha)^2 = \langle \delta_\alpha^2 \rangle_{\text{dis}} - \langle \delta_\alpha \rangle_{\text{dis}}^2, \tag{S41}$$

where

$$\langle \delta_\alpha^2 \rangle_{\text{dis}} = -\frac{1}{4N^4} \int dx\, dy\, dx'\, dy'\, \text{sign}(x-y)\,\text{sign}(x'-y')\, e^{2ik(x-y)} e^{2ik(x'-y')} \times$$

$$\times \left\langle \hat{n}_g(x)\hat{n}_g(y)\hat{n}_g(x')\hat{n}_g(y') e^{-i\frac{\pi\alpha}{N}(F(x)-F(y)+F(x')-F(y'))} \right\rangle_{\text{dis}}. \tag{S42}$$

Similar to previous case where we studied the fluctuations of $w$, the only non-vanishing contribution after applying Isserlis' theorem in the limit $k\sigma \gg \pi|\alpha|$ comes from the term with the pairing

$$\langle \delta\hat{n}_g(x)\delta\hat{n}_g(y')\rangle_{\text{dis}} \langle \delta\hat{n}_g(y)\delta\hat{n}_g(x')\rangle_{\text{dis}} = \left(\delta(x-y)n(x) - \frac{1}{N}n(x)n(y')\right)\left(\delta(y-x')n(y) - \frac{1}{N}n(y)n(x')\right). \quad \text{(S43)}$$

From these four terms, only the term $\delta(x-y')\delta(y-x')n(x)n(y)$ survives in the limit of interest and we are left with

$$\langle \delta_\alpha^2 \rangle_{\text{dis}} = -\frac{1}{4N^4} \int dx dy \, \text{sign}(x-y)\text{sign}(y-x)n(x)n(y) = \frac{1}{4N^4}\left(\int dx \, n(x)\right)^2 = \frac{1}{4N^2}. \quad \text{(S44)}$$

Thus, the fluctuations of $\delta_\alpha$ are given by

$$(\Delta\delta_\alpha)^2 \sim \frac{1}{N^2} \quad \text{(S45)}$$

and hence vanish faster than those of $w_\alpha$.

### Spectrum in perturbation theory

As shown above, the states $|\alpha, \text{F}\rangle$ and $|\alpha, \text{B}\rangle$ have vanishing overlap in the limit $k\sigma \gg \pi|\alpha|$ on average while the fluctuations of this quantity behave as $1/N$ in the limit $N \to \infty$. Further, we know that the matrix element $\delta_\alpha$ also vanishes on average while having fluctuations on the order of $1/N^2$. In the derivation of the universal dynamics in Sec. , we have shown that the dominant contribution comes from those states with small values of $\alpha$ where the overlap to the state $|W\rangle$ is large. Hence, we restrict our analysis to those values of $\alpha$ and aim at calculating the eigenenergies and -states of the full Hamiltonian in perturbation theory in order to derive some analytic formula for $P(t)$ in the presence of disorder.

For the perturbation theory, we assume that the unperturbed basis is given by $\{|\alpha, \text{FB}\rangle\}$ and the perturbation is given by both $H_\text{F}$ and $H_\text{B}$. This can be justified by noting that, for example, the matrix element $\delta_\alpha$ vanishes on average and has only small fluctuations for small $|\alpha|$ and large $N$. The same applies for the matrix element $\langle \alpha, \text{B}| H_\text{F} |\alpha, \text{B}\rangle$ and all off-diagonal elements. Since we are only interested in leading order effects in $N$, we apply degenerate perturbation theory in the subspace $\{|\alpha, \text{F}\rangle, |\alpha, \text{B}\rangle\}$ for fixed $\alpha$. Note, however, that these states are not orthogonal in general but only on average with fluctuations on the order of $1/N$ such that we have to orthogonalize this subspace artificially using

$$|\psi_1\rangle = |\alpha, \text{F}\rangle \quad \text{(S46)}$$

$$|\psi_2\rangle = \frac{1}{\sqrt{1-|w_\alpha|^2}}\left(|\alpha, \text{B}\rangle - w_\alpha^* |\alpha, \text{F}\rangle\right). \quad \text{(S47)}$$

In order to simplify the calculations, we neglect $\delta_\alpha$ since its fluctuations are much smaller than those of $w_\alpha$. The corresponding Hamiltonian in the subspace spanned by $|\psi_1\rangle$ and $|\psi_2\rangle$ then reads

$$H = E_\alpha \begin{pmatrix} 1 & \frac{w_\alpha^*}{\sqrt{1-|w_\alpha|^2}} \\ \frac{w_\alpha}{\sqrt{1-|w_\alpha|^2}} & \frac{1-3|w_\alpha|^2}{1-|w_\alpha|^2} \end{pmatrix}. \quad \text{(S48)}$$

Its eigenvalues are given by

$$E_\pm = \frac{E_\alpha}{1-|w_\alpha|^2}\left(1 - 2|w_\alpha|^2 \pm |w_\alpha|\right) \approx E_\alpha(1 \pm |w_\alpha|) \quad \text{(S49)}$$

to leading order in $w_\alpha$. The corresponding eigenstates are to leading order in $w_\alpha$ given by

$$|\alpha, +\rangle = \frac{1}{\sqrt{2}}\left(|\psi_1\rangle + \frac{w_\alpha^*}{|w_\alpha|}|\psi_2\rangle\right) \approx \frac{1}{\sqrt{2}}\left(|\alpha, \text{F}\rangle\left(1 - \frac{w_\alpha^{*2}}{|w_\alpha|}\right) + \frac{w_\alpha^*}{|w_\alpha|}|\alpha, \text{B}\rangle\right), \quad \text{(S50)}$$

$$|\alpha, -\rangle = \frac{1}{\sqrt{2}}\left(|\psi_1\rangle - \frac{w_\alpha^*}{|w_\alpha|}|\psi_2\rangle\right) \approx \frac{1}{\sqrt{2}}\left(|\alpha, \text{F}\rangle\left(1 + \frac{w_\alpha^{*2}}{|w_\alpha|}\right) - \frac{w_\alpha^*}{|w_\alpha|}|\alpha, \text{B}\rangle\right). \quad \text{(S51)}$$

In conclusion, the degenerate eigenenergies $E_\alpha$ split to first order in $w_\alpha$ with the splitting

$$\frac{\Delta E_\alpha}{E_\alpha} = 2|w_\alpha| \sim \frac{d_\alpha}{\sqrt{N}} \tag{S52}$$

where $d_\alpha$ is of order unity and depends on each single realization of the experiment. These fluctuating energies then set some time scale for the dephasing

$$\tau_{\rm dp} \sim \frac{\hbar}{\Delta E_\alpha} \sim \frac{1}{\sqrt{N}\gamma}. \tag{S53}$$

In terms of the time scale $\tau^{-1} \sim N\gamma$ given by the universal dynamics, we have

$$\tau_{dp} \sim \sqrt{N}\tau \tag{S54}$$

such that, relative to $\tau$, $\tau_{\rm dp}$ even grows with the particle number and makes the universal dynamics persisting for longer times.

## ANALYTIC EXPRESSION OF $P(t)$ IN THE LARGE $N$ LIMIT

Using the results calculated above, we are no able to derive analytically an approximate expression for $P(t)$ in the limit of large $N$ and $k\sigma \gg 1$. We start by writing

$$\begin{aligned}
P(t) &= \left|\langle W| e^{-\frac{i}{\hbar}Ht} |W\rangle\right|^2 \\
&\approx \left|\sum_\alpha e^{-i\frac{E_\alpha}{\hbar}t(1+|w_\alpha|)} |\langle W|\alpha,+\rangle|^2 + e^{-i\frac{E_\alpha}{\hbar}t(1-|w_\alpha|)} |\langle W|\alpha,-\rangle|^2 \right|^2 \\
&= \left|\sum_\alpha e^{-i\frac{E_\alpha}{\hbar}t} \left(e^{-i\frac{E_\alpha|w_\alpha|}{\hbar}t} |\langle W|\alpha,+\rangle|^2 + e^{i\frac{E_\alpha|w_\alpha|}{\hbar}t} |\langle W|\alpha,-\rangle|^2\right)\right|^2 \\
&= \sum_{\alpha\neq\beta} e^{-i\frac{E_\alpha}{\hbar}t} e^{i\frac{E_\beta}{\hbar}t} \left(e^{-i\frac{E_\alpha|w_\alpha|}{\hbar}t}|\langle W|\alpha,+\rangle|^2 + e^{i\frac{E_\alpha|w_\alpha|}{\hbar}t}|\langle W|\alpha,-\rangle|^2\right) \left(e^{-i\frac{E_\beta|w_\beta|}{\hbar}t}|\langle W|\beta,+\rangle|^2 + e^{i\frac{E_\beta|w_\beta|}{\hbar}t}|\langle W|\beta,-\rangle|^2\right) \\
&\quad + \sum_\alpha \left|e^{-i\frac{E_\alpha|w_\alpha|}{\hbar}t}|\langle W|\alpha,+\rangle|^2 + e^{i\frac{E_\alpha|w_\alpha|}{\hbar}t}|\langle W|\alpha,-\rangle|^2\right|^2.
\end{aligned} \tag{S55}$$

The overlap of the eigenstates $|\alpha,\pm\rangle$ with $|W\rangle$ is given by

$$|\langle W|\alpha,\pm\rangle|^2 = \frac{1}{2}\left[\left(\frac{2}{\pi\alpha}\right)^2 + |\langle W|\psi_2\rangle|^2 \mp \frac{2}{\pi\alpha}\left(e^{-i\phi_\alpha}\langle W|\psi_2\rangle + e^{i\phi_\alpha}\langle\psi_2|W\rangle\right)\right], \tag{S56}$$

where we set $w_\alpha = |w_\alpha|e^{i\phi_\alpha}$. Further, we have

$$\langle W|\psi_2\rangle = \langle W|\alpha,{\rm B}\rangle + |w_\alpha|e^{-i\phi_\alpha}\frac{2}{\pi\alpha} \tag{S57}$$

and

$$|\langle W|\psi_2\rangle|^2 = |\langle W|\alpha,{\rm B}\rangle|^2 + |w_\alpha|^2\left(\frac{2}{\pi\alpha}\right)^2 + \frac{2}{\pi\alpha}2|w_\alpha|\cos(\phi_\alpha). \tag{S58}$$

The overlap of the state $|W\rangle$ and the state $|\alpha,{\rm B}\rangle$ is given by

$$\langle W|\alpha,{\rm B}\rangle = \frac{1}{N}\int dx\, e^{-2ikx} \langle G|\hat{n}_g(x)e^{i\frac{\pi\alpha}{N}F(x)}|G\rangle \tag{S59}$$

and has the same distribution as $w_\alpha^*$ in the limits we are interested in. Thus, to leading order in $w_\alpha$, we can write

$$|\langle W|\alpha,\pm\rangle|^2 = \frac{1}{2}\left[\left(\frac{2}{\pi\alpha}\right)^2 + \frac{2}{\pi\alpha}|w_\alpha|2\cos(\phi_\alpha) \mp \frac{2}{\pi\alpha}|w_\alpha|\left(1+\frac{2}{\pi\alpha}\right)2\cos(2\phi_\alpha)\right]. \tag{S60}$$

Let us now focus on the second term in eq. (S55), where the sum runs only over $\alpha$. We can write

$$\sum_\alpha \left| e^{-i\frac{E_\alpha |w_\alpha|}{\hbar}t} |\langle W|\alpha,+\rangle|^2 + e^{i\frac{E_\alpha |w_\alpha|}{\hbar}t} |\langle W|\alpha,-\rangle|^2 \right|^2 = \sum_\alpha |\langle W|\alpha,+\rangle|^4 + |\langle W|\alpha,-\rangle|^4$$
$$+ 2|\langle W|\alpha,+\rangle|^2 |\langle W|\alpha,-\rangle|^2 \cos\left(\frac{2E_\alpha |w_\alpha|}{\hbar}t\right). \quad (S61)$$

Next, we insert the result obtained in eq. (S60) and neglect all terms of order $w_\alpha^2$. For the linear terms, we want to point out that they are all accompanied by a factor $\cos(\phi_\alpha)$ or $\cos(2\phi_\alpha)$. We verified numerically, that the complex random variable $w_\alpha$ follows a Gaussian distribution which is spherically symmetric such that $|w_\alpha|$ and $\phi_\alpha$ are independent random variables. The random phase $\phi_\alpha$ is uniformly distributed in the interval $[-\pi,\pi]$ and hence the cosine terms vanish after averaging over different realizations. Thus, we are left with

$$\sum_\alpha |\langle W|\alpha,+\rangle|^4 + |\langle W|\alpha,-\rangle|^4 + 2|\langle W|\alpha,+\rangle|^2 |\langle W|\alpha,-\rangle|^2 \cos\left(\frac{2E_\alpha |w_\alpha|}{\hbar}t\right) = \frac{16}{4\pi^4} \sum_\alpha \frac{2}{\alpha^4}\left(1 + \cos\left(\frac{2E_\alpha |w_\alpha|}{\hbar}t\right)\right). \quad (S62)$$

In addition, the averaging over $|w_\alpha|$ can be carried out by noting that $|w_\alpha|^2$ follows an exponential distribution with mean $1/N$ such that

$$\overline{\frac{16}{4\pi^4} \sum_\alpha \frac{2}{\alpha^4}\left(1 + \cos\left(\frac{2E_\alpha |w_\alpha|}{\hbar}t\right)\right)} = \frac{16}{2\pi^4} \sum_\alpha \frac{1}{\alpha^4} \int_0^\infty d|w_\alpha|^2 N e^{-N|w_\alpha|^2}\left(1 + \cos\left(\frac{2E_\alpha \sqrt{|w_\alpha|^2}}{\hbar}t\right)\right)$$
$$= \frac{16}{\pi^4} \sum_\alpha \frac{1}{\alpha^4}\left(1 - \frac{E_\alpha t\, D\left(\frac{E_\alpha t}{\hbar\sqrt{N}}\right)}{\hbar\sqrt{N}}\right), \quad (S63)$$

with the Dawson function $D(x)$ as defined below eq. (S34).

In a similar way, we can treat the first term in eq. (S55) which leads to

$$\overline{\sum_{\alpha \neq \beta} e^{-i\frac{E_\alpha}{\hbar}t} e^{i\frac{E_\beta}{\hbar}t}\left(e^{-i\frac{E_\alpha |w_\alpha|}{\hbar}t}|\langle W|\alpha,+\rangle|^2 + e^{i\frac{E_\alpha |w_\alpha|}{\hbar}t}|\langle W|\alpha,-\rangle|^2\right)\left(e^{-i\frac{E_\beta |w_\beta|}{\hbar}t}|\langle W|\beta,+\rangle|^2 + e^{i\frac{E_\beta |w_\beta|}{\hbar}t}|\langle W|\beta,-\rangle|^2\right)}$$
$$= \frac{16}{\pi^4} \sum_{\alpha \neq \beta} e^{-i\frac{E_\alpha}{\hbar}t} e^{i\frac{E_\beta}{\hbar}t} \frac{1}{\alpha^2}\frac{1}{\beta^2}\left(1 - \frac{E_\alpha t D\left(\frac{E_\alpha t}{2\hbar\sqrt{N}}\right)}{\hbar\sqrt{N}}\right)\left(1 - \frac{E_\beta t D\left(\frac{E_\beta t}{2\hbar\sqrt{N}}\right)}{\hbar\sqrt{N}}\right). \quad (S64)$$

Since $E_{-\alpha} = -E_\alpha$, we can restrict the sum over positive $\alpha$ and then replace $\alpha = 2n+1$ where $n = 0,1,2,\ldots$ Note that now we let the sum run over infinitely many integers even though we only have a finite number of states for finite $N$. The contributions from large $n$, however, give only small contributions such that we can safely neglect them. Further, we introduce $\tau = \pi/N\gamma$ such that finally the disorder-averaged dynamics of the state $|W\rangle$ is given by

$$P(t) = \frac{16}{\pi^4} \sum_{n \neq m \geq 0} \frac{2}{(2n+1)^2}\frac{2}{(2m+1)^2} \cos\left(\frac{t/\tau}{2n+1}\right)\cos\left(\frac{t/\tau}{2m+1}\right)\left(1 - \frac{t/\tau_\mathrm{dp}\, D\left(\frac{t/2\tau_\mathrm{dp}}{2n+1}\right)}{2n+1}\right)\left(1 - \frac{t/\tau_\mathrm{dp}\, D\left(\frac{t/2\tau_\mathrm{dp}}{2m+1}\right)}{2m+1}\right)$$
$$- \frac{16}{\pi^4} \sum_n \frac{2}{(2n+1)^4}\left[\left(1 - \frac{t/\tau_\mathrm{dp}\, D\left(\frac{t/2\tau_\mathrm{dp}}{2n+1}\right)}{2n+1}\right)^2 - \left(1 - \frac{t/\tau_\mathrm{dp}\, D\left(\frac{t/\tau_\mathrm{dp}}{2n+1}\right)}{2n+1}\right)\right]. \quad (S65)$$

In this expression, the first term is reminiscent of the universal dynamics showing oscillatory behavior which then is damped out on a time scale $\tau_\mathrm{dp} = \sqrt{N}\tau$, while for times $t \gg \tau_\mathrm{dp}$, the last term reduces to

$$\frac{16}{\pi^4} \sum_n \frac{2}{(2n+1)^4}\left(1 - \frac{t/\tau_\mathrm{dp}\, D\left(\frac{t/\tau_\mathrm{dp}}{2n+1}\right)}{2n+1}\right) \to \frac{16}{\pi^4} \sum_{n=0}^\infty \frac{1}{(2n+1)^4} = \frac{1}{6} \quad (S66)$$

which gives the universal saturation value different from zero. In Figure S1, we compare the dynamics calculated by numerically diagonalizing the Hamiltonian $H = H_\mathrm{F} + H_\mathrm{B}$ for $N=1000$ particles with a Gaussian density distribution satisfying $k\sigma = 100$ averaged over $10^5$ realizations (blue curve) with the expression (S65) obtained using perturbation theory (orange line) and find excellent agreement.

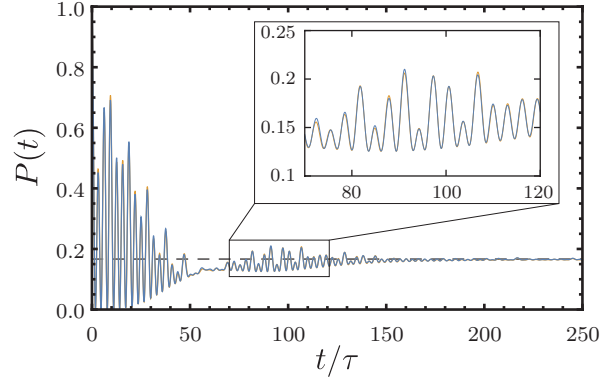

Figure S1. Comparison between the numerically and analytically calculated time evolution of $|W\rangle$ under $H_s$. Blue curve: Numerically calculated time evolution for $N = 1000$ particles averaged over $10^5$ realizations with a Gaussian distribution and $k\sigma = 100$. Orange curve: Analytically calculated curve using degenerate perturbation theory. (Inset:) Zoomed image of the dynamics for better comparison between numerically and analytically calculated time evolution.

## DISSIPATIVE DYNAMICS

The spontaneous emission rate into the backward propagating mode, $\Gamma_\text{B}$, depends on each experimental realization and is given by

$$\Gamma_\text{B} = \frac{\gamma}{N} \int dx dy \, e^{2ik(x-y)} \langle G| \hat{n}_g(x)\hat{n}_g(y) |G\rangle . \tag{S67}$$

Its average value can be obtained using eq. (S23) and as a result, we get

$$\bar{\Gamma}_\text{B} = \gamma(N-1)\frac{|n_{2k}|^2}{N^2} + \gamma, \tag{S68}$$

where $n_k$ is the Fourier transform of the density distribution $n(x)$. For a smooth density distribution with variations large compared to the optical wavelength $\lambda$, this contribution is strongl suppressed. For example focusing on a Gaussian distribution with width $\sigma$, $n(x) = Ne^{-x^2/2\sigma^2}/\sqrt{2\pi\sigma^2}$, we get $n_{2k}/N = e^{-8\pi^2\sigma^2\lambda^2}$ which is exponentially suppressed for $\lambda < \sigma$. Thus, the back scattering is not enhanced and is dominated by the incoherent contribution $\gamma$.

---


\end{document}